# MINIMIZING THE RISK OF ARCHITECTURAL DECAY BY USING ARCHITECTURE-CENTRIC EVOLUTION PROCESS

Humaira Farid, Farooque Azam and *M. Aqeel Iqbal

Department of Computer Engineering
College of Electrical & Mechanical Engineering
National University of Sciences and Technology (NUST), Islamabad, Pakistan
*[Employed at Faculty of E & IT, FUIEMS, Rawalpindi, Pakistan]

`humaira.farid@gmail.com, farooque.azam@gmail.com`
`maqeeliqbal@hotmail.com`

*ABSTRACT*

*Software systems endure many noteworthy changes throughout their life-cycle in order to follow the evolution of the problem domains. Generally, the software system architecture cannot follow the rapid evolution of a problem domain which results in the discrepancies between the implemented and designed architecture. Software architecture illustrates a system's structure and global properties and consequently determines not only how the system should be constructed but also leads its evolution. Architecture plays an important role to ensure that a system satisfies its business and mission goals during implementation and evolution. However, the capabilities of the designed architecture may possibly be lost when the implementation does not conform to the designed architecture. Such a loss of consistency causes the risk of architectural decay. The architectural decay can be avoided if architectural changes are made as early as possible. The paper presents the Process Model for Architecture-Centric Evolution which improves the quality of software systems through maintaining consistency between designed architecture and implementation. It also increases architecture awareness of developers which assists in minimizing the risk of architectural decay. In the proposed approach consistency checks are performed before and after the change implementation.*

*KEYWORDS*

*Software Architecture, Software Evolution, Architectural Decay, Architecture Versioning, Architecture Assessment*

## 1. INTRODUCTION

Software systems are usually designed to provide a solution to a particular problem domain and for a particular business case. As the business world is frequently changing and the problem domains evolve the software systems have to be constantly tailored to new business needs, i.e. they need to evolve [1].
Software evolution activities can be classified as to correct errors that are found in operation (corrective), to adapt it for a new platform (adaptive) and to improve its performance by adding new functionality or other non-functional characteristics (perfective).

Software architecture illustrates a system's structure and global properties and consequently determines not only how the system should be constructed but also leads its evolution. The stability is an important criterion for evaluating the architecture. The stability of the architecture is a measure of how well it accommodates the evolution of the system without requiring changes to the architecture. Consider the Figure-1 which shows the distribution of evolution effort.

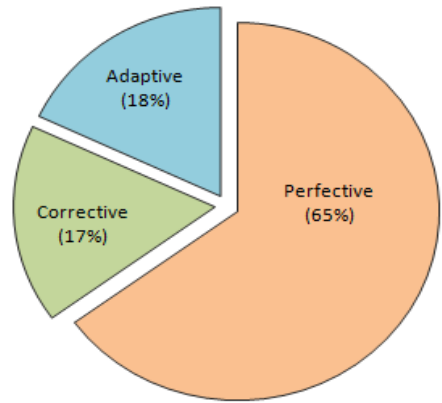

Figure-1: Distribution of Evolution Effort

Architectural stability is more vulnerable by changes in non-functional rather than in functional requirements [2]. Figure-2 shows that the architecture plays an important role to ensure that a system satisfies its business and mission goals during implementation and evolution.

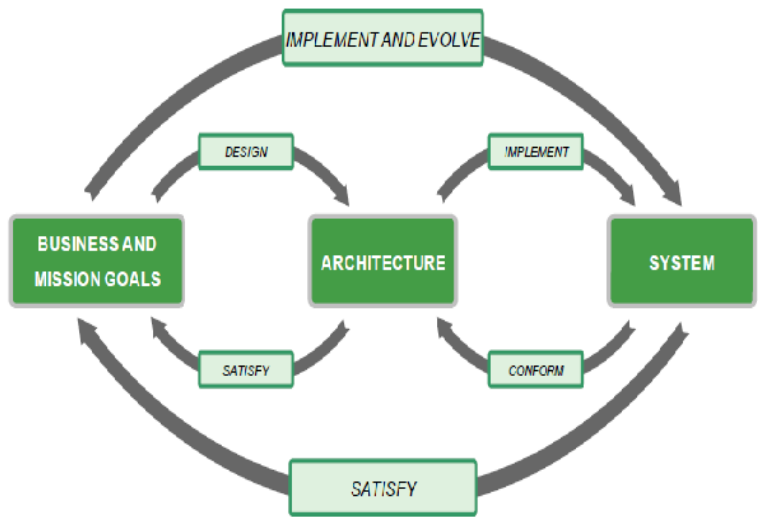

Figure-2: Architecture-Centric Development & Evolution

However, the capabilities of the designed architecture may possibly be lost when the implementation does not conform to the designed architecture. Such a loss of consistency causes the risk of architectural decay. The paper describes that architecture plays an important role in improving software quality and provides a solid basis for software evolution. The paper emphasizes on the importance of early and rapid architecture evolution for minimizing the risk of architectural decay.

The paper presents the Process Model for Architecture-Centric Evolution which improves the quality of software systems through maintaining consistency between designed architecture and implementation. It also increases architecture awareness of developers which assists in minimizing the risk of Architectural Decay. In the proposed approach consistency checks are performed before and after the change implementation. It evaluates the implemented architecture for identifying the risk of architecture's quality decay and inconsistencies between the architecture and the implementation. The rest of this paper is organized as follows. Section II describes the problem description. Section III presents the proposed architecture-centric evolution process model and also defines every process and its sub-activities in detail. Section IV illustrates application areas and Section V defines potential research areas. Finally, Section VI concludes the paper.

## 2. SCOPE OF RESEARCH

Software systems endure many noteworthy changes throughout their life-cycle in order to follow the evolution of the problem domains. Generally, the software system architecture cannot follow the rapid evolution of a problem domain which results in the discrepancies between the implemented and designed architecture [3]. The preferred way of working is to add new system features in an ad-hoc manner without changing the architectural description which results in architectural decay and degrades the overall software quality.

Architecture refactoring is required to avoid this problem. Generally, Architecture refactoring is deferred until the very last moment when it becomes extremely essential. Delays in performing small refactoring activities turn into need for architecture reengineering which is more risky and expensive. The effectiveness of reengineering is also generally not as high as anticipated [1]. Usually maintenance pays attention on comparatively small changes due to time and budget limitations without considering structural changes, which can lead to imperfect changes and consequent errors.

As a countermeasure it is good practice to maintain the consistency between the architecture and the implementation. Consistency checking can be done by deducing information from implementation, design documents, and model transformations [4]. Design decisions made at the architectural level directly affect system maintenance and evolution. Hence, a considerable effort is spent on designing architecture to assist future evolution. However, this effort may possibly be lost, if the implementation deviates from the designed architecture. Such divergence between the design and implementation results in Architectural Decay which makes further maintenance tasks more complex and expensive [5].

## 3. PROPOSED APPROACH

The architectural decay can be avoided if architectural changes are made as early as possible. This paper introduces the Architecture-Centric Evolution Process Model, which supports keeping system architecture up-to-date with the problem domain and thus minimizing the risk of architectural decay and quality degradation. The model introduces the evolution life cycle in which it integrates the Architecture evolution with code evolution. It supports maintaining consistency between architecture and implementation and thus offering the solid basis for effective evolution.

The process model includes the four fundamental activities (see figure-3); Evolution Analysis and Validation, Architecture Evolution, Change Implementation and Architecture Assessment.

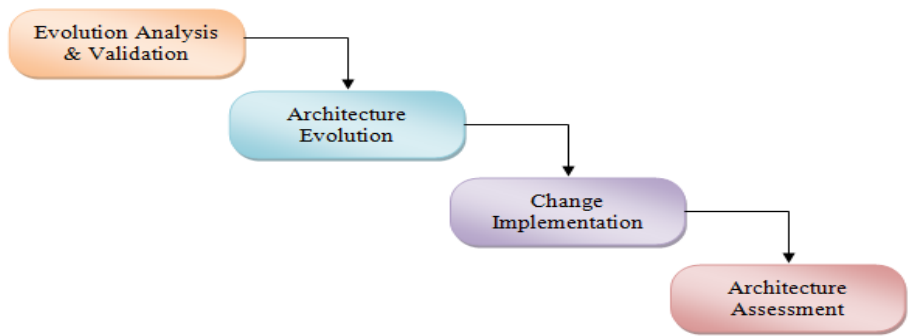

Figure-3: Architecture-Centric Evolution Process Model

The Evolution Analysis and Validation stage of this model examines the impact of change and checks its consistency in architecture description. It also analyses and validates the new requirements that reflect the system changes. If changes cause the risk of architectural decay, the risk severity and urgency of proposed changes will be analyzed. Architecture Evolution modifies the architecture description according to requested changes. If inconsistencies are detected, the evolution process considers the proposed evolution as being part of a new architecture version. Change Implementation modifies the system specification and implements it in the implementation environment to reflect the changes. The Architecture Assessment stage evaluates the implemented architecture for identifying the risk of architectural decay and inconsistencies between the architecture and the implementation.

Every process takes some input for performing its task. After the completion of task, it gives some output which assists next process for achieving its task. Figure-4 shows the key inputs and outputs of every process in evolution life cycle.

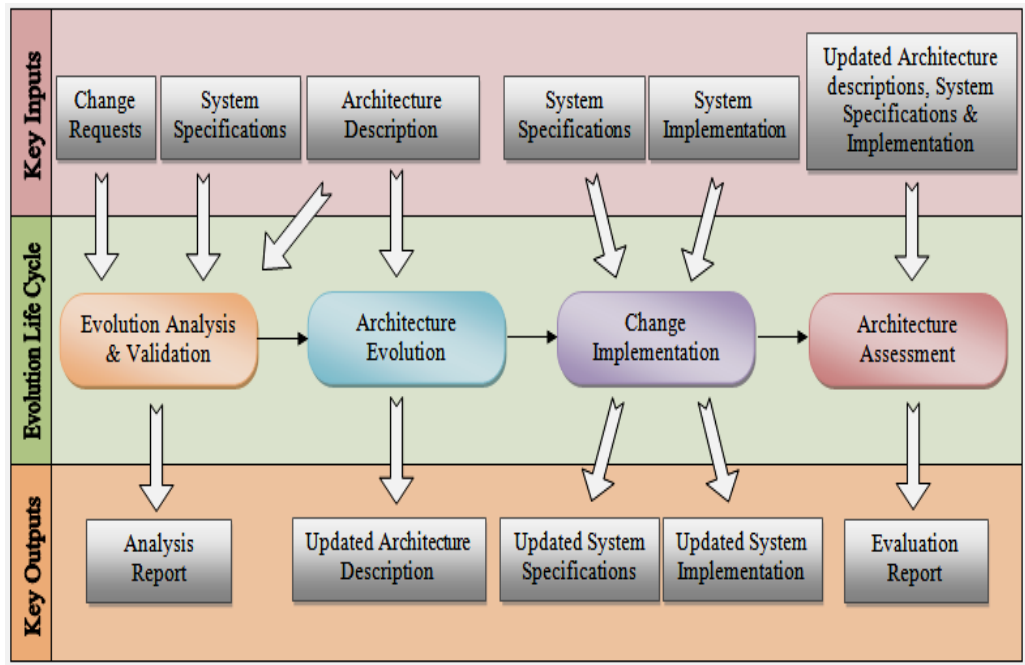

Figure-4: Architecture-Centric Evolution Process Model with Key inputs & outputs

4International Journal of Computer Science, Engineering and Applications (IJCSEA) Vol.1, No.5, October 2011

Every process in evolution life cycle has some sub-activities. Each sub-activity performs a particular task and provides output for assisting other activities. Table-1 illustrates the sub-activities of every process along with key inputs and outputs.

Table-1: Evolution Processes with Sub-Activities and Key inputs & outputs

| Process | Sub-Activities | Key Inputs | Key Outputs |
| --- | --- | --- | --- |
| Evolution Analysis & Validation | Change Impact Analysis | Change Requests | Change Plan Analysis Report |
| | Requirement Validation | System Specifications | |
| | Consistency Checking | Architecture Descriptions | |
| Architecture Evolution | Architecture Modification | Current Architecture Description | Updated Architecture Descriptions |
| | Architecture Versioning | | |
| Change Implementation | System Specification Updating | Current System Specifications | Updated System Specifications & Implementation |
| | Source Code Modification | Current System Implementation | |
| Architecture Assessment | Architecture Assessment | Updated Architecture Descriptions Updated System Specifications & Implementation | Evaluation Report |

The rest of the section is organized to explain every process along with its sub-activities in detail.

## 3.1 Evolution Analysis and Validation

Evolution Analysis and Validation examines the impact of change and checks its consistency in architecture description. It also analyses and validates the new requirements that reflect the system changes. It includes three sub-activities.

- Change Impact Analysis
- Requirement Validation
- Consistency Checking

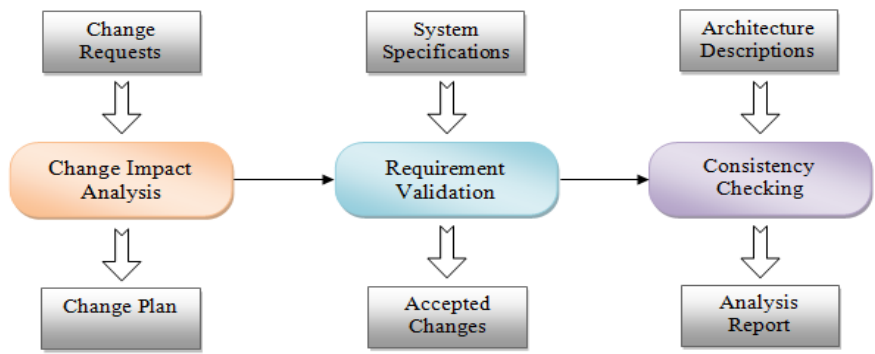

Figure-5: Evolution Analysis & Validation Process

Change Impact Analysis takes the input of requested changes. The impact of these changes is evaluated to see how much of the system is affected by the change and how much it might cost to implement the change. If the proposed changes are accepted, a new release of the system is planned. The outcome of this activity is the release of Change Plan which highlights the places where changes are required in Architecture descriptions, system specifications and implementation. New requirements that reflect the system changes are proposed, analyzed and validated in Requirement Validation activity. The requirements are analyzed in detail to check whether these requirements satisfy the business goals. The proposed changes are only accepted if they do not contradict the business goals.

After Requirement Validation, architectural consistency will be ensured. Consistency Checking uses change plan and current architecture descriptions to check whether architectural elements will contradict one another after implementing the proposed changes. The consistency checking aims to predict whether changes persuade inconsistencies in current architecture. If changes maintain consistency, the proposed evolution will be permitted. If not, it will either be unacceptable or trigger the derivation of a new architecture version for which consistency will be guaranteed. The outcome of this activity is the Analysis Report.

If changes cause the risk of architectural decay, the risk severity and urgency of the proposed changes will be analyzed. For this assessment Analysis Matrix is introduced. Table-2 shows the matrix which illustrates the different categories.

Table-2: Analysis Matrix for the assessment of the Risk of Architectural Decay along with the urgency of requested changes.

| Urgency of Change | Severity of the Risk of Architectural Decay | | | |
|---|---|---|---|---|
| | Minor A | Moderate B | Critical C | Catastrophic D |
| 1-Routine | 1A | 1B | 1C | 1D |
| 2-Urgent | 2A | 2B | 2C | 2D |
| 3-Most Urgent | 3A | 3B | 3C | 3D |

After assessment of the risk of architectural decay and urgency of the proposed changes, it will require to perform some necessary measures for ensuring quality and consistency. Table-3 shows the required measures against every category.

Table-3: Assessment index and corresponding required measures for maintaining consistency and improving quality.

| Assessment Index | Measures |
| --- | --- |
| 1A, 1B, 1C, 1D, 2C, 2D, 3D | First Architecture will be updated and afterwards the implementation will be modified. |
| 3C | A decision will be taken according to Business & missions goals. |
| 2A, 2B, 3A, 3B | First Implementation will modify and immediately afterwards Architecture evolution will take place. |

The details of the above assessment along with their corresponding required measures will be included in the Analysis Report.

### 3.2 Architecture Evolution

Architecture Evolution modifies the architecture description according to requested changes. It includes two sub-Activities:

- Architecture Modification
- Architecture Versioning

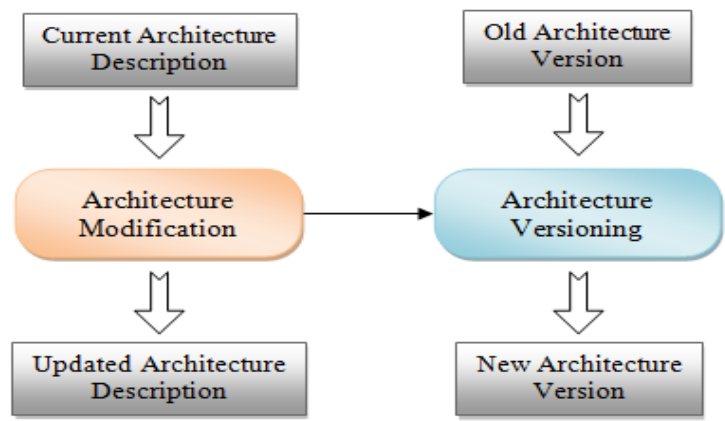

Figure-6: Architecture Evolution Process

Architecture descriptions will be modified according to the Change Plan and the Analysis report. This activity takes input of current architecture description to make changes. Architecture modification is used to modify the architectural description according to the required changes introduced in the analysis report. It assists in maintaining consistency between system architecture and implementation. The updated architecture description is a major outcome of this activity.

If inconsistencies are detected, the evolution process considers the proposed evolution as being part of a new architecture version. Architecture Versioning derives new architecture version for consistency with the proposed change and maintains all previous versions. It also records all important change operations performed in the previous architecture version.

### 3.3 Change Implementation

Change Implementation modifies the system specification and implements it in the implementation environment to reflect the changes. It includes two sub-activities:

- System Specification Updating
- Source Code Modification

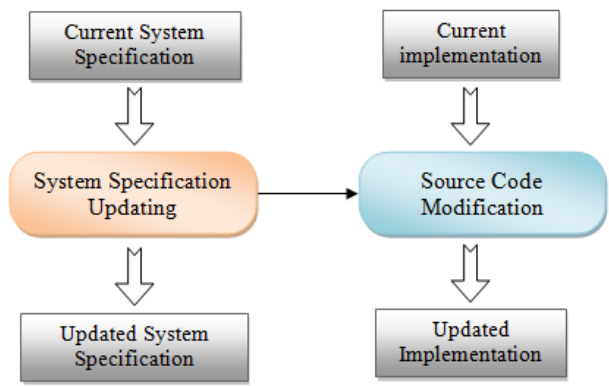

Figure-7: Change Implementation Process

System Specification Updating takes input of current system specification for updating. The system specification is modified according to new requirements mentioned in change plan. The updated System Specification is a key outcome of this activity.

In Source code modification proposed changes are implemented in the implementation environment to reflect the changes. It takes the input of current implementation and updates it according to updated system specification and architecture description.

### 3.4 Architecture Assessment

Architecture Assessment aims to evaluate architecture after it has been implemented. Assessment of an implemented architecture assists in identifying the risk of architecture's quality decay and inconsistencies between the architecture and the implementation.

Architecture Assessment takes inputs of updated architecture descriptions, system specifications and implementation for checking consistency. The outcome of this phase is an evaluation report containing the results of the assessment and specific actions for adjustment in case of any discrepancies between the architecture and implementation.

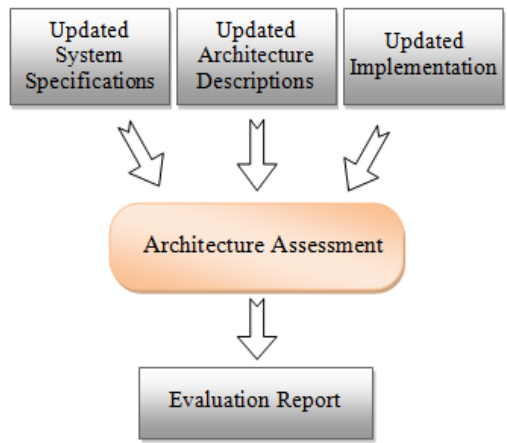

Figure-8: Architecture Assessment Process

Figure-8 shows the key inputs Architecture Assessment activity is taking for achieving its task. It also shows a key outcome of this activity i.e. an Evaluation Report.

## 4. APPLICATION AREAS

### 4.1 Evolution in product lines and families

Software product families are a set of independent programs that have several common and variable functionalities. Software product lines have received extensive adoption in many software companies. A wide variety of companies has significantly reduced the cost of software development and maintenance and improved the quality of their software products. The product line approach can be applied to an existing line of products or the organization can also use a new system or product family to expand its market. In case of adding new products, product line architecture and components need to evolve with the requirements posed by new product line members. Architecture-centric evolution process assists in evolving the software product line architecture. It keeps architecture consistent and improves overall quality.

### 4.2 Evolution of legacy software through its architecture

The legacy software systems are described as old software systems which are usually designed and documented inadequately, but still perform an important job for the business critical application. The business value of legacy systems has become feeble due to the lack of consistency and evolution support. But the importance of legacy systems cannot be undermined as some of their functions are too important to be scrapped completely and too costly to reconstruct. Organizations have to make a pragmatic assessment of legacy systems to choose the most suitable approach for evolving these systems. Architecture-centric evolution of legacy systems improves their business value by providing consistency and improving overall quality.

### 4.3 Evolution of EAI services-oriented architecture

Information systems are now based on integration of existing components that have to cooperate in a precise manner in order to build a services-based application. The EAI (Enterprise Application Integration) domain provides integration models and techniques for assembling various software applications in a realistic way. EAI architecture defines the elements that compose the system and their interaction. The evolution support and the inconsistency between design and implementation are the major issues addressed by designing and building COTS-based systems [6]. To maintain the consistency between the architecture and the implementation, architecture-centric evolution approach is used.

### 4.4 Architecture-centric evolution process for component-based software

Component-based software engineering (CBSE) emerged as a reuse-based approach to software development. It promotes an approach to define, implement and integrate or compose loosely coupled independent components into systems [7]. A new component role can be required to add to cope with new requirements. The specification of software architecture will also be required to evolve to meet new requirements. Architecture-centric evolution process provides a controlled support for component-based software evolution that prevents architecture drift and erosion [8].

## 5. POTENTIAL RESEARCH AREAS

### 5.1 Architecture-centric evolution process for modern development methodologies: RAD, Agile and Extreme Programming

Rapid Application Development (RAD) is a modern software development methodology that uses nominal planning in support of rapid prototyping. There can be real difficulties with this approach. Without a specification it may be difficult to validate the system.
Frequent changes have a tendency to corrupt software structure and it makes it more expensive to change for meeting new requirements. The integration of Agile approaches and software architecture is possible but it requires that professionals from both fields work together to overcome evident challenges in this field and should emphasize on the need of research on integrating these two paradigms. [9]

### 5.2 Developing new metrics and approaches supporting Architecture-centric evolution process

Different metrics and patterns can be applied for the software evolution management. New metrics and approaches could be used in the architecture-centric evolution process for assuring the quality of a software system not only in the software design phase but also throughout the software development life cycle. This could be done by calculating a variety of design metrics from the system architecture and reporting prospective quality harms to the designers and developers. This could assist in improving the software quality and minimizing the risk of architectural decay.
### 5.3 Tools that maintain and impose Architecture-centric evolution process

The automatic tool support for Architecture-centric evolution process could make it more effective and less time consuming. Consistency checking and architecture assessment could be done effectively and easily by using efficient tools. The tool support could be provided for automatically detecting the architectural changes and apply them in the implementation environment. This could minimize the time and effort for software evolution.

## 6. CONCLUSION

This paper has described that architecture plays an important role in improving software quality and provides a solid basis for software evolution. The paper has emphasized on the importance of early and rapid architecture evolution for minimizing the risk of architectural decay. This paper has defined that how software architecture illustrates a system's structure and global properties and leads the system evolution. This paper has proposed the Process Model for Architecture-Centric Evolution for improving the quality of software systems through maintaining consistency between the architecture and implementation. The paper has argued that the proposed approach increases architecture awareness of developers which assists in minimizing the risk of Architectural Decay. In the proposed approach consistency check has been performed before and after the change implementation. The proposed process model includes the four fundamental activities; Evolution Analysis and Validation, Architecture Evolution, Change Implementation and Architecture Assessment. The Evolution Analysis and Validation stage of this model examines the impact of change and checks its consistency in architecture description. It also analyses and validates the new requirements that reflect the system changes. Architecture Evolution modifies the architecture description according to requested changes. Change Implementation modifies the system specification and implements it in the implementation environment to reflect the changes. Finally, the Architecture Assessment stage evaluates the implemented architecture which assists in identifying the risk of architecture's quality decay and inconsistencies between the architecture and the implementation.

## REFERENCES

[1]. Ilian Pashov, "Feature-Based Methodology for Supporting Architecture Refactoring and Maintenance of Long-Life Software Systems", PhD thesis, TU Ilmenau.
[2]. Rami Bahsoon and Wolfgang Emmerich, "Architectural Stability and Middleware: An Architecture-Centric Evolution Perspective", workshop on Architecture-Centric Evolution 2006.
[3]. "Architecture-Centric Evolution: New Issues and Trends", Report on the Workshop ACE at ECOOP-2006
[4]. Matthias Biehl and Welf Löwe, "Automated Architecture Consistency Checking for Model Driven Software Development", Proceedings of the 5th International Conference on the Quality of Software Architectures: Architectures for Adaptive Software Systems, 2009.
[5]. Jacek Rosik, Jim Buckley, Muhammad Ali Babar, "Design Requirements for an Architecture Consistency Tool", in Proceedings of the 21st Annual Psychology of Programming Interest Group Conference, June 2009.
[6]. Frédéric Pourraz, Hervé Verjus, "An architecture-centric approach for managing the evolution of EAI services-oriented Architecture", in Proceedings of the Eighth International Conference on Enterprise Information Systems Databases and Information Systems Integration, 2006.
[7]. I. Sommerville, "Software Engineering", 8th ed. Addison Wesley, 2006.
[8]. HY Zhang, Christelle Urtado, Sylvain Vauttier, "Architecture-centric development and evolution processes for component-based software", in Proceedings of SEKE'2010.
[9]. Babar Ali M., and A. Pekka, "Architecture-Centric method and agile approaches", 10th international conference on agile processes, Dec, 2009.

**Humaira Farid**

Humaira Farid is a student of M.S (Computer Software Engineering) in the department of Computer Engineering, College of Electrical and Mechanical Engineering, National University of Sciences and Technology (NUST), Pakistan.

**Dr. Farooque Azam**

Dr. Farooque Azam is an Associate Professor in the Department of Computer Engineering, College of Electrical and Mechanical Engineering, National University of Sciences and Technology (NUST), Pakistan. He did his PhD in Software Engineering from BUAA, Beijing, China. He did his MS in Software Engineering from Military College of Signals, National University of Sciences and Technology (NUST), Pakistan. 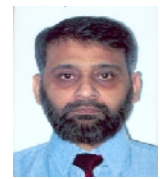

**M. Aqeel Iqbal**

M. Aqeel Iqbal Is An Assistant Professor In The Department Of Software Engineering, Faculty Of Engineering And Information Technology, Foundation University, Institute Of Engineering And Management Sciences, Rawalpindi, Pakistan. As A Researcher He Has A Deep Affiliation With The College of E & ME, National University Of Sciences And Technology (NUST), Pakistan.

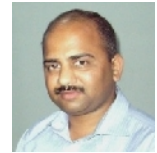